\RequirePackage{fix-cm}
\documentclass[twocolumn]{svjour3}          
\smartqed  
\usepackage{graphicx}
\usepackage{marvosym}
\journalname{Preprint}
\begin{document}

\title{Electron propagation from a photo-excited surface: implications for time-resolved photoemission
}


\author{S.-L. Yang         \and
        J. A. Sobota          \and
        P. S. Kirchmann      \and
        Z.-X. Shen 
}


\institute{S.-L. Yang \and J. A. Sobota \and Z.-X. Shen \at
             Stanford Institute for Materials and Energy Sciences, SLAC National Accelerator Laboratory, 2575 Sand Hill Road, Menlo Park, CA 94025, USA\\
             Geballe Laboratory for Advanced Materials, Departments of Physics and Applied Physics, Stanford University, Stanford, CA 94305, USA\\
\\
           P. S. Kirchmann(\Letter) \at Stanford Institute for Materials and Energy Sciences, SLAC National Accelerator Laboratory, 2575 Sand Hill Road, Menlo Park, CA 94025, USA\\
                Tel.: +1-650-926-3522\\
                Fax: +1-650-926-5191\\
                \email{kirchman@slac.stanford.edu}
}

\date{Received: 11 October 2013/ Accepted: 8 November 2013}

\maketitle

\begin{abstract}
We perform time- and angle-resolved photoelectron spectroscopy on p-type GaAs(110). We observe an optically excited population in the conduction band, from which the time scales of intraband relaxation and surface photovoltage decay are both extracted. Moreover, the photovoltage shift of the valence band intriguingly persists for hundreds of picoseconds at negative delays. By comparing to a recent theoretical study, we reveal that the negative-delay dynamics reflects the interaction of the photoelectrons with a photovoltage-induced electric field outside the sample surface. We develop a conceptual framework to disentangle the intrinsic electron dynamics from this long-range field effect, which sets the foundation for understanding time-resolved photoemission experiments on a broad range of materials in which poor electronic screening leads to surface photovoltage. Finally, we demonstrate how the long-lasting negative-delay dynamics in GaAs can be utilized to conveniently establish the temporal overlap of pump and probe pulses in a time-resolved photoemission setup.

\keywords{photoelectron spectroscopy \and ultrafast dynamics \and surface photovoltage \and semiconductors}
\end{abstract}

\section{Introduction}
\label{intro}
Time- and angle-resolved photoelectron spectroscopy (trARPES) has emerged as a powerful technique to study non-equilibrium properties of quantum materials \cite{Schmitt2008,Sobota2012,Wang2012,Hajlaoui2012,Perfetti2007,Cortes2011,Graf2011,Smallwood2012,Rettig2013}. In this technique, an infrared (IR) pump pulse first excites the sample, and an ultraviolet (UV) probe pulse subsequently photoemits electrons to unveil the transient population and energy dynamics. The access to electronic band structure upon optical excitation enables a detailed investigation into the evolution of photo-excited charge carriers, which is directly connected to various quantum phenomena such as melting of the charge-density wave order \cite{Schmitt2008}, spin-polarized transport in topological insulators \cite{Sobota2012,Wang2012,Hajlaoui2012}, and quasiparticle dynamics in unconventional superconductors \cite{Perfetti2007,Cortes2011,Graf2011,Smallwood2012,Rettig2013}. In this respect, a thorough understanding of the pump-probe photoemission process is of fundamental importance to the successful application of trARPES to these condensed matter physics problems.

It is usually conceived that trARPES reflects the transient spectral function $A(\bf{k},\omega,\tau)$ \cite{Schmitt2008}, where $\bf{k}$, $\omega$, and $\tau$ stand for momentum, energy, and pump-probe delay, respectively. For $\tau<0$, the sample is regarded as being in thermal equilibrium, and no dynamics are to be expected. However, this concept relies on the assumption that the energy and momentum of a photoelectron are completely determined at the moment of photoemission, and its propagation in vacuum is not affected by the electron dynamics in the sample. The latter assumption is violated if the electron dynamics lead to a time-dependent electric field near the sample surface, which can occur in materials with spatially separated electron-hole pairs \cite{Rettig2012,Muraoka2004,Schmuttenmaer1996,Siffalovic2002,Tokudomi2008,Azuma2009,Widdra2003,Lim2005,Tanaka2012}. The effect of this transient field on electron propagation has been observed \cite{Widdra2003,Lim2005}, but it is not clear how to disentangle the intrinsic electron dynamics from the transient field effect.

In this work we investigate a simple model system, p-type GaAs(110), and its surface photovoltage (SPV) effect. The photo-excited electron-hole pairs are separated by the electric field associated with surface band-bending. This separation subsequently generates a long-range dipole field outside the sample which impacts the propagation of photoelectrons in vacuum. Our trARPES results demonstrate that the SPV-generated field gives rise to a valence band (VB) energy shift persisting for hundreds of picoseconds (ps) at negative delays, yet the SPV decay time constant of $\sim$1.5 ps is still correctly resolved for both the valence and the conduction bands (CB) at positive delays. We explain this behavior by considering the electrodynamics along the electron trajectory. Our results not only agree well with a recent theoretical study \cite{Tanaka2012}, but more importantly define the appropriate concepts to understand trARPES data when such transient field effects are substantial.

\section{Experimental setup}

Our trARPES setup is based on a 80 MHz Ti-Sapphire oscillator system. We refer to Ref. \cite{Sobota2012} for an extensive discussion of the experimental details. The 1.5 eV, 50-femtosecond (fs) IR beam is split into pump and probe paths. In the pump path, an optical delay stage provides a tunable pump-probe delay. In the probe path, 6 eV UV photons are obtained by frequency-quadrupling. Transient photoelectron spectra are obtained with a hemispherical analyzer upon UV probing. The combined temporal resolution is measured to be $\sim$160 fs. All the pump beam radii are referenced to half-width-half-maximum $r_{0}\sim57$ $\mu$m. The incident IR fluence used in this work is 25 $\mu$J/cm$^{2}$ unless specified otherwise, and the measurements were done at room temperature. We used Zn-doped p-type GaAs(110) samples from MTI corporation with the doping level of $1.2\times 10^{19}$ cm$^{-3}$. For this doping level, the bulk Fermi level (E$_{F}$) is expected to be 0.03 eV above the VB maximum, and 1.39 eV below the CB minimum \cite{Sze2007}. The samples were cleaved \emph{in situ} under ultra-high vacuum (UHV) with a pressure $<1\times10^{-10}$ Torr.

As shown in the inset of Fig. \ref{Fig1}(c), we define a positive pump-probe delay $\tau$ as the configuration in which the IR pump pulse is incident on the sample before the UV probe pulse, and a negative delay as the configuration in which the probe pulse is incident before the pump pulse. Time zero refers to exact temporal overlap of the two pulses.

\section{Results and discussion}

\begin{figure}  
\begin{center}
\includegraphics[width=\columnwidth]{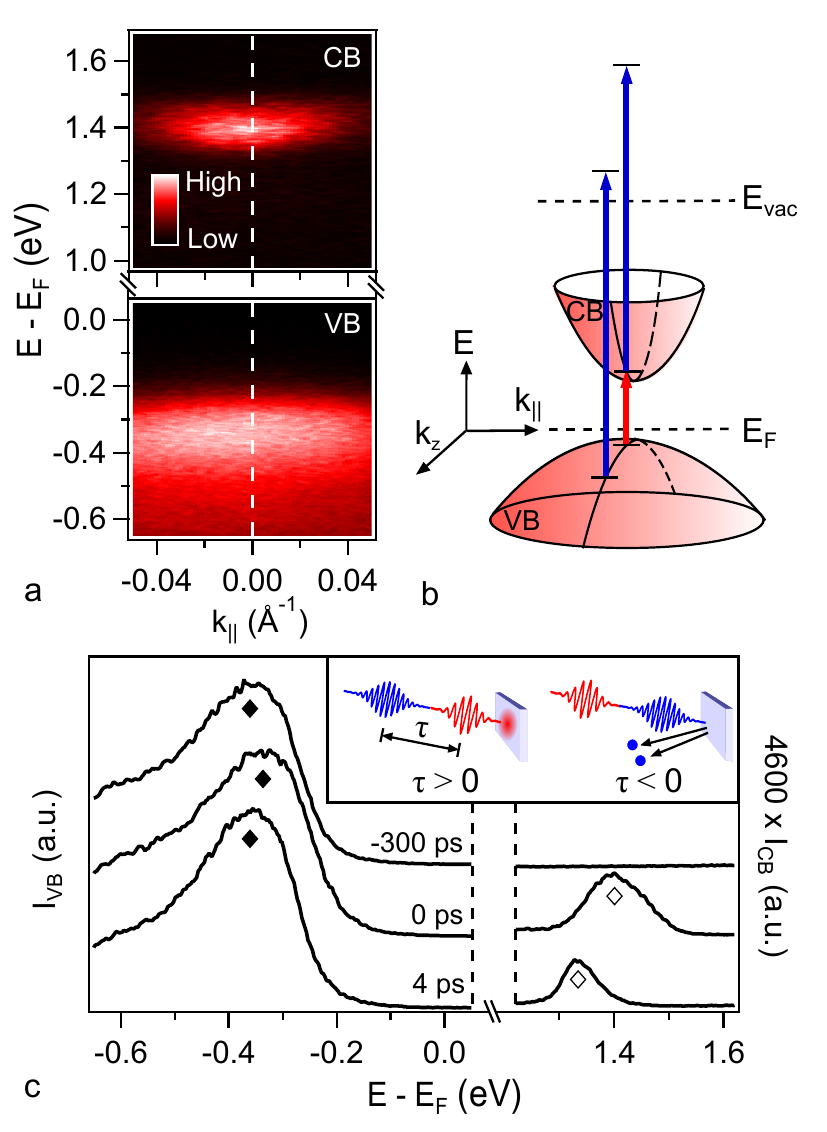} 
\caption{\small (color online). Identification of spectral features within the GaAs band structure. (a) Photoelectron spectrum for both CB and VB near the $\bar{\Gamma}$-point at time zero. (b) Schematic band structure near E$_{F}$ and optical transitions induced by 1.5 and 6 eV photons (red and blue arrows). (c) EDCs at $\bar{\Gamma}$ for -300, 0, and 4 ps. Solid and hollow diamonds in (c) indicate the VB and CB peak positions. Inset: illustration of the configurations for positive and negative delays.} \label{Fig1}
\end{center}
\end{figure}  

We present the trARPES spectrum at time zero in Fig. \ref{Fig1}(a). Due to optical excitation of electrons across the band gap, we resolve an electron population in the CB. The CB and VB features are observed near 1.40 eV and -0.33 eV with respect to E$_{F}$. While the CB energy is close to the expected bulk CB minimum, the VB energy is significantly different from the expected bulk VB maximum. We attribute the VB shift mainly to the downward band-bending due to defect-induced surface states \cite{Gudat1976,Mele1979,Chelikowsky1979}, along with 6 eV photons probing a non-zero out-of-plane momentum (k$_{z}$) within the hole-like VB dispersion shown in Fig. \ref{Fig1}(b) \cite{Chiang1980}. In contrast, a non-zero k$_{z}$ would shift the CB peak upward because of the electron-like CB dispersion, and hence compensate its downward shifting due to band-bending.

We plot selected energy distribution curves (EDCs) in Fig. \ref{Fig1}(c) to illustrate the delay-dependent energy shifting. The spectra at time zero display an upward shift compared to those at -300 and 4 ps. As discussed in literature \cite{Siffalovic2002,Tokudomi2008,Azuma2009}, this transient shift is consistent with the SPV picture, in which the downward band-bending on a p-type semiconductor traps photo-excited electrons at the surface and transiently shifts the band structure up in energy. It is worth noting that previous pump-probe photoemission studies on the subject of GaAs SPV did not report the CB shifting dynamics \cite{Siffalovic2002,Tokudomi2008,Azuma2009}. The observation of both the CB and VB dynamics allows us to provide a complete account of the SPV physics, and address the important issue of how the SPV-induced field affects the interpretation of the trARPES data.

\begin{figure}  
\begin{center}
\includegraphics[width=\columnwidth]{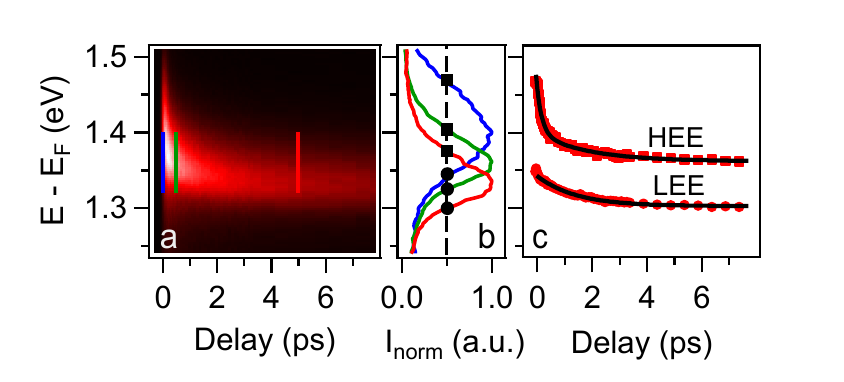} 
\caption{\small (color online). CB energy dynamics. (a) $\bar{\Gamma}$-point EDCs as a function of pump-probe delay. (b) Comparison of normalized EDCs at 0 (blue), 0.5 (green), and 5 (red) ps. Solid squares and circles represent the half-maximum points of the HEE and LEE. (c) Transient energy dynamics for the HEE and LEE of the CB spectral peak. The HEE is fitted by a double-exponential function: $\tau_{H1}=200\pm20$ fs, $\tau_{H2}=1.8\pm0.3$ ps. The LEE is fitted by a single-exponential function: $\tau_{L}=1.4\pm0.1$ ps.} \label{Fig2}
\end{center}
\end{figure}  

We proceed to discuss the CB energy dynamics in more detail by plotting EDCs at $\bar{\Gamma}$ as a function of pump-probe delay, as presented in Fig. \ref{Fig2}(a). Fig. \ref{Fig2}(b) compares normalized EDCs at 0, 0.5 and 5 ps. Besides the overall energy relaxation, there is a notable difference between the evolution of the higher-energy edge (HEE) and that of the lower-energy edge (LEE). We extract the half-maximum points to quantitatively determine the edge positions as a measure of the changing lineshape. In the first 0.5 ps, the HEE displays a significantly larger shift than the LEE, whereas after 0.5 ps, the HEE and LEE are shifted by nearly the same energy. 

We extract the time scales of the HEE and LEE decay dynamics in Fig. \ref{Fig2}(c). While the LEE decay follows a single-exponential function with the time constant of $\tau_L = 1.4\pm0.1$ ps, the HEE decay can only be captured by the sum of two exponential functions with the time constants of $\tau_{H1} = 200\pm20$ fs and $\tau_{H2} = 1.8\pm0.3$ ps.

Notably there are two distinct time scales in the CB energy dynamics ($\tau_{H1}=200$ fs, $\tau_{H2}\approx\tau_{L}\approx$1.5 ps). We point out that $\tau_{H1}$ is associated with the intraband relaxation of the CB electrons, which has been similarly observed in other semiconductors and attributed to electron-phonon interactions \cite{Sobota2012,Ichibayashi2009}. This interpretation is consistent with the observation that the LEE, which represents the bottom of our observed CB, does not exhibit the fast decay dynamics. On the other hand, the slower energy dynamics in the CB reflect the relaxation of SPV. The fact that the HEE and LEE exhibit the same time constant after 0.5 ps is consistent with the electrostatic nature of SPV: it rigidly shifts the whole band structure and leaves the CB peak width unchanged.

\begin{figure}  
\begin{center}
\includegraphics[width=\columnwidth]{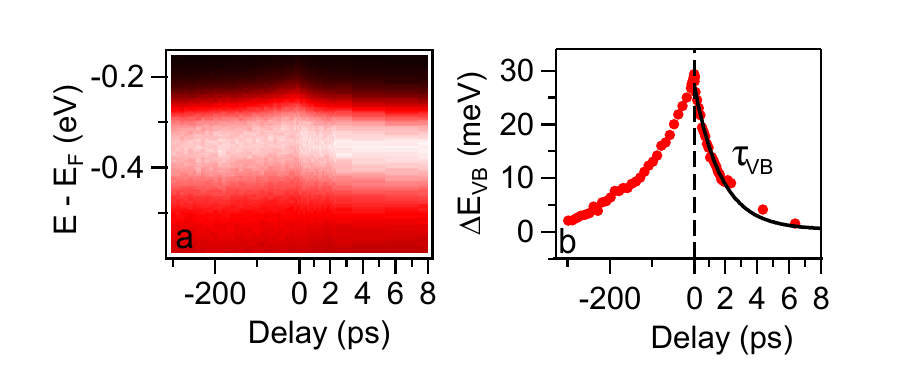} 
\caption{\small (color online). VB energy dynamics. (a) $\bar{\Gamma}$-EDCs as a function of pump-probe delay. Note the horizontal axis scaling is different for positive and negative delays. (b) Extracted VB energy dynamics. For positive delays, the transient energy shift decays with a time constant of $\tau_{VB}=$ 1.7 $\pm$ 0.1 ps.} \label{Fig3}
\end{center}
\end{figure} 

This SPV picture is further substantiated by the VB dynamics, which is shown in Fig. \ref{Fig3}. In Fig. \ref{Fig3}(a) we plot $\bar{\Gamma}$-EDCs as a function of pump-probe delay. We extract the delay-dependent VB energy as displayed in Fig. \ref{Fig3}(b). For positive delays, the VB energy decays exponentially with $\tau_{VB}=$ 1.7 $\pm$ 0.1 ps, which agrees well with the larger time constant of the CB dynamics. This similarity indicates that the two processes share the same origin, which again points to SPV since it is expected to rigidly shift the whole band structure.

\begin{figure*}  
\begin{center}
\includegraphics[width=1.7\columnwidth]{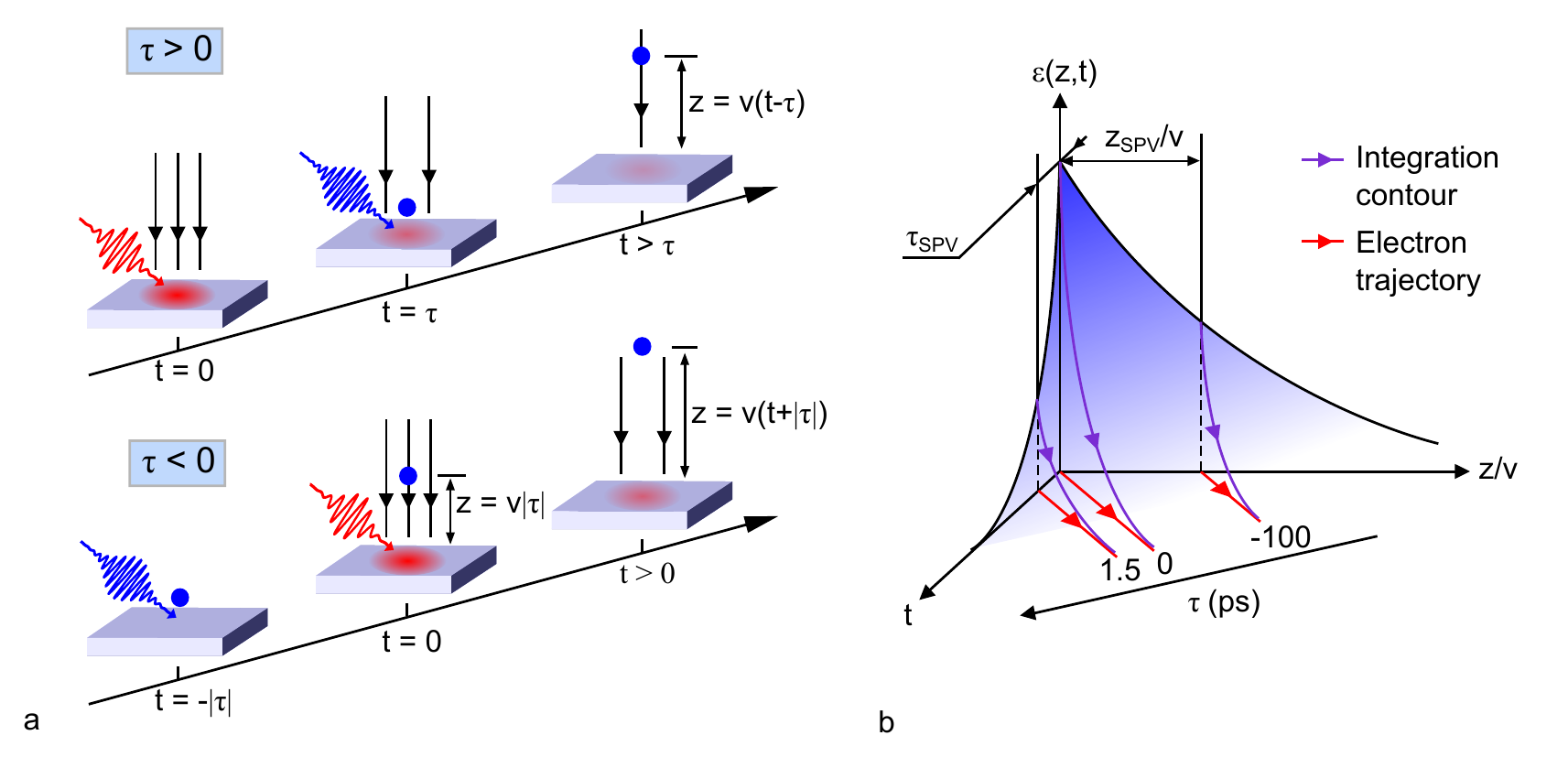} 
\caption{\small (color online). Interpretation of the positive- and negative-delay dynamics in VB. (a) Comparison of the event sequences. The black lines perpendicular to the sample surface represent the SPV-induced electric field. (b) Comparison of the time scales. The purple lines refer to the integration contours, and the red lines represent the approximate electron trajectories in the $z$-$t$ plane.} \label{Fig4}
\end{center}
\end{figure*}

Intriguingly, the VB energy is not constant before time zero, which is what we would expect since the sample is in equilibrium before excitation. Instead, it continues shifting to lower energies at negative delays on a time scale of $\sim$100 ps. A recent theoretical study by Tanaka predicted this phenomenon based on the fact that a photoemitted electron is not instantaneously measured by the detector, but must propagate for a finite time in the electric field generated by SPV \cite{Tanaka2012}. This interaction between the IR-induced electric field and the photoelectrons in vacuum not only concerns the photoemission spectra at negative delays, but also indicates that the positive-delay spectra are not momentary snapshots of the transient band structure, since the electron propagation time is usually much longer than the ultrafast electron dynamics \cite{Tanaka2012}. Therefore, the electron propagation needs to be carefully considered when interpreting trARPES data. Inspired by Tanaka's theoretical investigation \cite{Tanaka2012}, we construct a simple conceptual framework to help disentangle different physical processes.

We consider the sequence of events for positive- and negative-delay configurations in Fig. \ref{Fig4}(a). For positive delays, the IR pulse first induces the SPV and its corresponding electric field. The field affects the propagation of photoelectrons during their travel. For negative delays, however, the photoelectrons first travel to a distance $v|\tau|$ away from the surface, and then start to be affected by the IR-induced field. We define $t$ as the time referenced to the moment of pump incidence, and $\tau$ as the pump-probe delay. Considering that the electron trajectory for $t\geq\tau$ is approximately $z_{\tau}(t)=v\cdot (t-\tau)$, where $v$ is the speed of the electron, we may write down the general expressions for the energy changes in the positive- and negative-delay configurations.

\begin{equation}\label{final}
\Delta E(\tau) = \left\{\begin{array}{ll} -e\int_{0}^{\infty}\varepsilon(z,z/v+\tau)dz\,\,\,&\tau\geq 0\\
    -e\int_{v|\tau|}^{\infty}\varepsilon(z,z/v-|\tau|)dz\,\,\,&\tau<0
       \end{array} \right.
\end{equation}

Here $\varepsilon(z,t)$ is the electric field as a function of space and time. 

Without performing the integrals, we can gain insights from a graphical representation of the integration. In Fig. \ref{Fig4}(b) we plot the schematic electric field $\varepsilon(z,t)$. The red lines in the $z$-$t$ plane represent the electron trajectories for different delays. The purple lines represent the corresponding integration contours on the $\varepsilon(z,t)$ surface. 

In the positive-delay configuration, the integration in Eqn. \ref{final} always goes from $z=0$ to $z=\infty$. As the delay $\tau$ increases, the integration contour shown in Fig. \ref{Fig4}(b) moves along the $\varepsilon(z=0,t)$ curve. Using $\varepsilon(z,t)\sim\xi(z)\exp{(-t/\tau_{SPV})}$ as a realistic ansatz, the integration result for positive delays will be proportional to $\exp{(-\tau/\tau_{SPV})}$. This is indeed one of the important results of Ref. \cite{Tanaka2012}, and explains why we observe the SPV decay time constant in the CB and VB for positive delays.

For negative delays, we may change variables to the time $t$.

\begin{equation}\label{negdelay}
\Delta E(\tau)=-ev\int_{0}^{\infty}\varepsilon(v(t+|\tau|),t)dt
\end{equation}

Analogous to the situation for positive delays, the integration contour moves along the $\varepsilon(z,t=0)$ curve as $|\tau|$ increases. In other words, different negative delays correspond to photoelectrons sensing the field at different distances from the sample surface. In fact, as $\tau_{SPV}$ is negligible compared to the negative-delay time scale, the photoelectrons in vacuum simply undergo an impulsive acceleration following the field at distance $v|\tau|$. Therefore, the response of photoelectrons exactly depicts the spatial profile of the field. The characteristic time scale at negative delays is proportional to the characteristic length scale $z_{SPV}$, which is determined by the beam radius $r$. 

\begin{figure}  
\begin{center}
\includegraphics[width=\columnwidth]{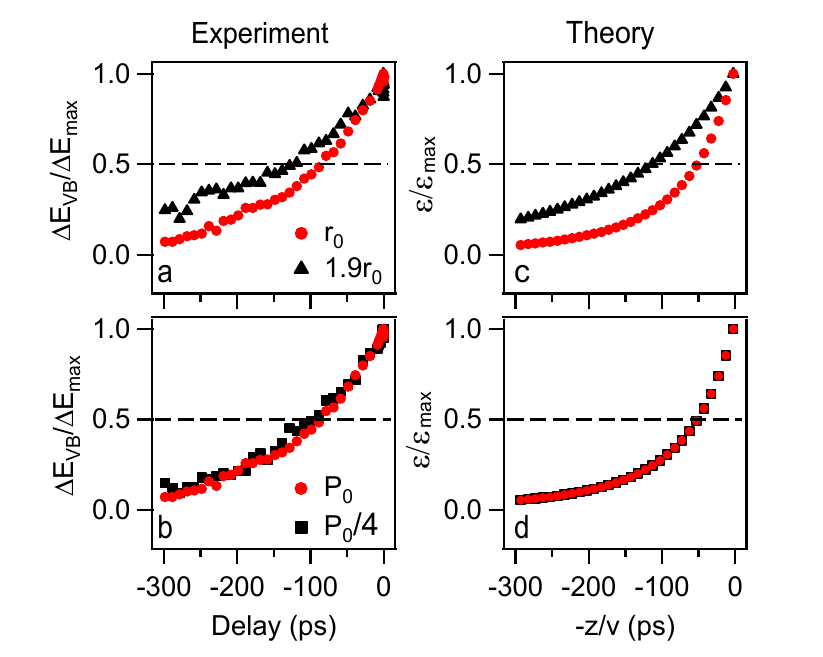} 
\caption{\small (color online). Comparison between experiment and theory on the negative-delay dynamics. (a) Normalized VB energy shift at negative delays for different IR beam radii while keeping the pulse energy constant. (b) Normalized VB energy shift at negative delays for different IR pulse energies while keeping the beam radius constant. (c, d) Calculated spatial profiles of the IR-induced dipole field corresponding to the configurations in (a) and (b), respectively.} 
\label{Fig5}
\end{center}
\end{figure}

We can use this concept to model the negative-delay dynamics. In Fig. \ref{Fig5}(a) we compare the negative-delay dynamics for two different beam radii ($r_{0}$, 1.9$r_{0}$) and fixed IR pulse energy ($P_{0}$). The shift of the half-maximum point indicates a change in the negative-delay time scale. As a control experiment, we also compare in Fig. \ref{Fig5}(b) the negative-delay dynamics for two IR pulse energies ($P_{0}$, $P_{0}$/4) and fixed radius ($r_{0}$). As the IR fluence is proportional to $P/r^{2}$, the two data sets in Fig. \ref{Fig5}(b) are associated with the same fluences as those in (a). The negative-delay time scale is demonstrated in (b) to be independent of the IR fluence, so the time scale difference in (a) is attributed to the change in beam radius. 

Fig. \ref{Fig5}(c) and (d) plot the calculated spatial profiles of the SPV-generated field for the corresponding configurations in (a) and (b). Here we assume the dipole distribution ${\bf p}(x,y)$ on the surface to be proportional to the IR intensity profile, and perform the calculation as in Equation \ref{calc}.

\begin{equation}\label{calc}
\varepsilon (z) \sim \int dxdy\left(\frac{3({\bf p}(x,y)\cdot\hat{\bf r})\hat{\bf r}-{\bf p}(x,y)}{r^{3}}\right)\cdot\hat{{\bf z}}
\end{equation}

Here $r=\sqrt{x^{2}+y^{2}+z^{2}}$ is the distance between the infinitesimal source ${\bf p}(x,y)dxdy$ and the point at $(0,0,z)$. $\hat{\bf z}$ is the unit vector along the surface normal direction. The calculation agrees reasonably well with the experimental results and supports our conceptual understanding that the negative-delay time scale is determined by the characteristic length scale of the SPV-generated field. 

We continue by discussing the implications of this negative-delay phenomenon from a more general perspective. The pump-induced field effect can be a generic phenomenon for materials in which the photoexcited charge distribution gives rise to a macroscopic electric field. This would not be the case in metals, because excess charges are efficiently screened and the time-integrated Coulomb field effect is negligible. However, due to less efficient screening this field effect may be substantial for a wide variety of semiconductors including the recently discovered topological insulators. Our work defines a general framework to disentangle the intrinsic electron dynamics from the long-range field effect in trARPES experiments.

On the other hand, the negative-delay dynamics represents a convenient tool in the routine operation of trARPES experiments. To set up a trARPES measurement, the pump and probe beams must be spatially and temporally overlapped on the sample surface. The latter is particularly challenging, since hot electron lifetimes in metals are typically on the order of $\sim$100 fs \cite{Petek1997}, which corresponds to an optical path length of 30 $\mu$m. Since the pump and probe paths are typically $\sim$1 m long, establishing the temporal overlap to such precision can be a significant experimental challenge. The $\sim$100 ps time scale observed in GaAs alleviates this difficulty since it corresponds to a significantly longer path length of $\sim$3 cm. In addition, this effect is electrostatic, and persists even with the spatial overlap not perfectly optimized. Finally, no sophisticated sample preparation is necessary: we have confirmed that this effect exists in samples loaded into vacuum directly from atmosphere, without any {\it in-situ} cleavage or surface cleaning. Due to these advantages, we regularly utilize GaAs as a convenient tool for establishing temporal overlap in our trARPES experiment.

\section{Conclusion}

trARPES is emerging as a powerful tool to study electron dynamics directly in the energy, momentum and time domains. We use the ultrafast electron dynamics in p-type GaAs(110) as a simple model system to test the fundamental principles of this technique. We observe an optically-induced population in the CB, which exhibits two distinct energy relaxation time scales: one corresponds to the intraband relaxation while the other reflects the SPV decay. In addition, we report a negative-delay VB energy shift with a time constant of $\sim100$ ps. We explain this behavior based on the coupling of photoelectrons to the SPV-generated electric field, and find good agreement between theory and experiment. This phenomenon has significant implications in the research field of trARPES, and we develop the appropriate concepts to uncover the intrinsic electron dynamics dressed by the SPV-induced field effect. Finally, the long-lasting negative-delay dynamics in GaAs can be used as a tool for conveniently establishing the temperal overlap of pump and probe pulses, which greatly improves the efficiency of our trARPES experimental setup.

\begin{acknowledgements}
We thank Makoto Gonokami, Dan Riley, and Jared Schwede for stimulating discussions. J. A. S. acknowledges support by the Stanford Graduate Fellowship. This work is supported by the U.S. Department of Energy, Office of Basic Energy Sciences, Division of Materials Science.
\end{acknowledgements}

\bibliographystyle{spphys}       
\bibliography{GaAs_rev3_citation}

\end{document}